\documentclass[12pt, letter]{article}
\usepackage[T2A]{fontenc} 
\usepackage[utf8]{inputenc} 
\usepackage[english,russian]{babel} 
\usepackage{amsmath, amssymb}
\usepackage[english]{babel}
\usepackage{xcolor, graphicx}
\usepackage{multicol}
\usepackage[left=3cm, right=2cm, top=2cm, bottom=2cm]{geometry}
\usepackage[space]{cite}
\usepackage[displaymath]{lineno}

\DeclareMathOperator{\Wr}{Wr}
\newcommand{\pa}{\partial}
\newcommand{\la}{\lambda}
\newcommand{\opsi}{\overline{\psi}}
\newcommand{\ophi}{\overline{\phi}}

\renewcommand{\vec}{\mathbf}
\newcommand{\sech}{\mathop{\rm sech}\nolimits} %
\begin{document}

UDC 530.18 + 532.59 + 538.574

{\centering
\bigskip

{\Large\bf \phantom{Lump interactions with plane solitons}\\ Lump interactions with plane solitons\footnote{This paper was submitted to the special issue of the journal Radiophysics and Quantum Electronics dedicated to the 80-th Birthday of Professor M.I. Rabinovich on 20 April 2021.}}

\bigskip

\textit{Y.~A.~Stepanyants$^{1,2;}$\footnote{Corresponding author, Yury.Stepanyants@usq.edu.au},  D.~V.~Zakharov$^{3}$, and V.~E.~Zakharov$^{4,5}$}

\bigskip
\phantom{abc}

$^1$~School of Mathematics, Physics and Computing, University of Southern Queensland, 487--535 West Street, Toowoomba, Queensland, 4350, Australia;

$^2$~Department of Applied Mathematics, Nizhny Novgorod State Technical University n.a. R.E. Alekseev, Nizhny Novgorod, Russia;

$^3$~Department of Mathematics, Central Michigan University, USA;

$^4$~Lebedev Institute of the Russian Academy of Sciences, Moscow, Russia;

\hspace{2.5cm}$^5$~Skolkovo Institute of Science and Technology, Moscow, Russia.
}%

\mbox{}

\hspace{7cm} {\bf  Abstract}

\noindent We analyse in detail the interactions of two-dimensional solitary waves called lumps and one-dimensional line solitons within the framework of the Kadomtsev--Petviashvili equation describing wave processes in media with positive dispersion. We show that line solitons can emit or absorb lumps or periodic chains of lumps; they can interact with each other by means of lumps. Within a certain time interval, lumps or lump chains can emerge between two line solitons and disappear then due to absorption by one of the solitons. This phenomenon resembles the appearance of rogue waves in the oceans. The results obtained are graphically illustrated and can be applicable to the description of physical processes occurring in plasma, fluids, solids, nonlinear optical media, and other fields.

\newpage

\begin{multicols}{2}
\noindent Я, человек отжалевший науке \\
Лучшие годы, житель Земли, \\
Жду, чтобы взяли истории руки \\
Нас за грудки, да и всласть потрясли. \\
\\
\\
\\
I, a person who has given\\
the best of my years to science, \\
an Earthling, am waiting for \\
History's hands to grab us by the lapels \\
and give us a good shake. \\
\phantom{Zakharov, Zakharov} V.E. Zakharov \\
\end{multicols}

\section{INTRODUCTION}

Slightly more than half a century ago, in 1970 B.B. Kadomtsev and V.I. Petviashvili published a seminal paper where an equation generalising the well-known Korteweg--de Vries (KdV) equation for the two-dimensional case was derived \cite{Kadomtsev-1970}. 
This equation, later named the Kadomtsev--Petviashvili (KP) equation \cite{Zakharov-1974}, was designated for the description of nonlinear dispersive waves travelling primarily in one direction with a smooth variation in the perpendicular direction. 
In the dimensional form, the KP equation can be written as:
\begin{equation}
\label{KPeq-Dim}
    \frac{\partial}{\partial \xi}\left(\frac{\partial v}{\partial \tau} + c\,\frac{\partial v}{\partial \xi} + \alpha\,  v\,\frac{\partial v}{\partial \xi} + \beta\,\frac{\partial^3 v}{\partial \xi^3}\right) = -\frac{c}{2}\frac{\partial^2 v}{\partial \eta^2},
\end{equation}
where $v(\xi, \eta, \tau)$ is the variable describing a perturbation of a particular field (for example, water surface elevation or plasma density, etc.), $c$ is the speed of linear long waves, $\alpha$ and $\beta$ are the coefficients of nonlinearity and dispersion, respectively; they depend on the particular physical problem. This equation can be converted to the dimensionless form which is convenient for the further analysis:
\begin{equation}
\label{DimLesKPeq}
    \frac{\partial}{\partial x}\left(\frac{\partial u}{\partial t} + 6u\, \frac{\partial u}{\partial x} + \frac{\partial^3 u}{\partial x^3}\right) = -3\gamma\,\frac{\partial^2 u}{\partial y^2},
\end{equation}
where $u = \alpha v/6\beta$, $t = \beta \tau$, $x = \xi - c\tau$, $y = \eta\sqrt{6|\beta|/c}$, and $\gamma = \pm 1$ is the dispersive parameter that plays an important role in determining the physical and mathematical properties of solutions of this equation.

One of the simplest solutions of the KP equation is a plane soliton which can run at a small angle to the $x$-axis:
\begin{equation}
\label{KdVsol}
    u(x,y,t) = A \sech^2{\left(kx + ly - \omega t\right)},
\end{equation}
where $k$ and $l$ are arbitrary parameters, $A = 2 k^2$ is the soliton amplitude, $\omega = 4 k^3 + 3\gamma l^2/k$. The soliton velocity is: 
\begin{equation}
\label{Solvel}
    \vec V = \left(\frac{\omega k}{k^2 + l^2}, \frac{\omega l}{k^2 + l^2}\right), \quad V \equiv |\vec V| = \frac{4 k^4 + 3\gamma l^2}{k\sqrt{k^2 + l^2}},
\end{equation}
where $V$ is the soliton speed.

Note that in the 1D case when $l = 0$ and the KP equation reduces to the KdV equation, the soliton speed is $V_{KdV} = \omega/k = 4 k^2 = 2A$. In the 2D case assuming $l \ll k$, we obtain from Eq. (\ref{Solvel}):
\begin{equation}
\label{Solvel2}
    V = \frac{4 k^4 + 3\gamma l^2}{k\sqrt{k^2 + l^2}} \approx 4 k^2\left[1 - \frac 12\left(1 - \frac 32\frac{\gamma}{k^2} \right)\frac{l^2}{k^2}\right] = V_{KdV}\left[1 - \frac{1}{2}\tan^2{\varphi} \left(1 - \frac{6\gamma}{V_{KdV}}\right) \right],
\end{equation}
where $\varphi = \tan^{-1}{(l/k)}$ is the angle between the soliton velocity $\bf V$ and the $x$-axis. In accordance with the KP-approximation, the angle $\varphi$ must be small,  $\varphi \ll 1$, therefore $V \approx V_{KdV}$ up to the small correction $\sim \varphi^2$.

As was shown in the original paper \cite{Kadomtsev-1970}, plane solitons are stable with respect to small perturbations along their fronts only if the dispersion parameter $\gamma > 0$; otherwise they are unstable and experience a self-focusing instability if  $\gamma < 0$. In the former case, Eq. (\ref{DimLesKPeq}) is called the KP2 equation; it is {\it strongly integrable} in the terminology of Ref. \cite{Zakharov-2009}. In the latter case of $\gamma < 0$, Eq. (\ref{DimLesKPeq}) is called the KP1 equation; it is {\it weakly integrable} \cite{Zakharov-2009}. The development of the self-focusing instability of plane solitons leads to the creation of {\it lumps} -- completely localised two-dimensional solitary waves. 

\begin{figure}[!htbp]
\centering
\includegraphics[width=6.0in]{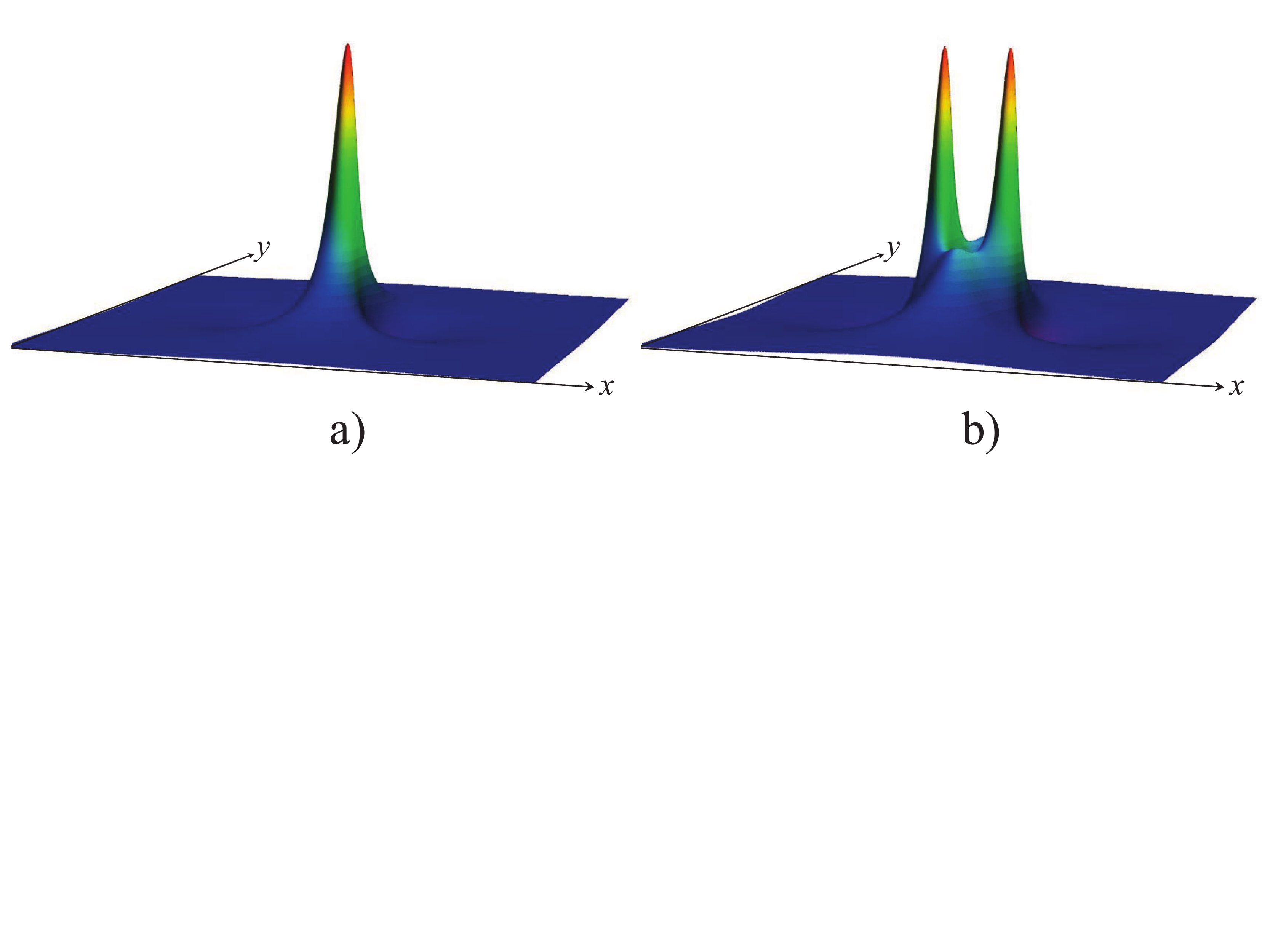}
\vspace{-6.0cm}%
\caption{Examples of a single lump (a) and bi-lump (b) in the KP1 equation.}
\label{f01}
\end{figure}

The lump solution to the KP1 equation was constructed numerically for the first time by V.I. Petviashvili in his seminal paper \cite{Petviashvili-1976} (see also \cite{Petviashvili-1992}), then such solutions were found analytically in Ref. \cite{Manakov-1977}. The solution to Eq. (\ref{DimLesKPeq}) with $\gamma = -1$ describes a symmetric lump moving along the $x$-axis is (see Fig. \ref{f01}a):
\begin{equation}
\label{Lump}
    u = 12V\frac{9 + V^2y^2 - 3V(x - Vt)^2}{\left[9 + V^2y^2 + 3V(x - Vt)^2\right]^2}\,,
\end{equation}
where $A = 4V/3$ is the lump amplitude, and $V > 0$ is the lump speed. There are more general solutions describing lumps travelling at arbitrary angles to the $x$-axis \cite{Manakov-1977, Lu-2004, {Singh-2016}}. Lumps interact elastically with each other \cite{Manakov-1977, Lu-2004, Hu-2018} not even undergoing phase shifts. Due to their non-monotonic asymptotics with the local minima, they can create stationary multi-lump formations \cite{Pelinovsky-1993A, Gorshkov-1993, Singh-2016, Hu-2018}. One of the simplest multi-lump formation, the bi-lump, is shown in Fig. \ref{f01}b (the details of solutions can be found in the cited papers).

A single lump is stable with respect to small perturbations \cite{Kuznetsov-1982}, whereas multi-lump structures and periodic chains of lumps are unstable \cite{Gorshkov-1993}.
In particular, a periodic lump chain under a small periodic perturbation along a chain front experiences decay into two new chains which are unstable in turn with respect to small perturbations and so on. 
This process repeats again and again with smaller and smaller decay rates \cite{Pelinovsky-1993B}. 
The lifetime of a rarefied chain of lumps with a long distance between them can be fairly big, therefore such chains can be observable in experiments and in nature.
Similarly, if the plane soliton has a periodic perturbation along its front, then as a result of the focusing instability, a new plane soliton of a smaller amplitude arises accompanied by the periodic chain of lumps moving side by side to each other \cite{Pelinovsky-1993B}.
Such an instability occurs only with respect to long perturbations along the $y$-axis, $\Lambda > \Lambda_c$, where $\Lambda_c$ is inverse proportional to the soliton amplitude and is much greater than the soliton width Ref. \cite{Zakharov-1975}.

As was recently shown \cite{Lester-2021}, lump chains moving at an angle to each other can interact experiencing spatial phase shifts similar to interacting plane solitons (see, e.g., \cite{Ablowitz-1981}). 
Under certain conditions, two interacting lump chains can form a single lump chain or vice versa, one lump chain can split into two lump chains moving in space at an angle to each other. The entire pattern moves stationary similar to a plane soliton triad under the resonance interaction within the KP2 equation (see \cite{Stepanyants-2020} and references therein). 

Another interesting observation was made in the recent publications \cite{Guo-2021, Rao-2021, Rao-2022}, where it was shown that within the KP1 equation, there are solutions that represent lump emission from a plane soliton or lump absorption  by a plane soliton. 
A lump can be emitted by one plane soliton and absorbed by another plane soliton. A lump can be obscured between two plane solitons travelling parallel to each other and having equal amplitudes at infinity. 
There are many other nice solutions representing a number of plane solitons exchanging lumps.

Here we review such solutions and analyse in detail the interaction of lumps with plane solitons using the Grammian form of the $\tau$-function \cite{Lester-2021}. We show that such an approach allows one to present different types of soliton and lump interactions in a relatively simple and natural way.

\section{INTERACTION OF A LINE SOLITON WITH LUMPS}
\label{Sect-2}

There are two standard formulae for the solution of the KP1 equation in terms of the $\tau$-function, both of them involving a set of solutions to an auxiliary linear system. Of the two, the more commonly used is known as the {\it Wronskian formula}:
\begin{equation}
\label{eq:Wronskian}
    u(x, y, t) = 2\,\frac{\partial^2 \ln{\tau}}{\partial x^2}, \quad \tau(x, y, t) = \Wr(\psi_1,\ldots,\psi_M).
\end{equation}
Here $\Wr(\psi_1,\ldots,\psi_M)$ is the Wronskian determinant of a linearly independent set of solutions of the system 
\begin{equation}
\psi_y=i\psi_{xx},\quad \psi_t=-4\psi_{xxx}.
\label{eq:linearsystem}
\end{equation}
Note that this is a set of linear equations which can be solved by separation of variables. 
This approach has been used in Ref. \cite{Johnson-1978} for the construction some simplest solutions within the general Zakharov--Shabat scheme \cite{Zakharov-1974}.

In our paper, we instead consider solutions defined by the so-called {\it Grammian formula}. As before, let $(\psi_1,\ldots,\psi_M)$ be a set of solutions (not necessarily linearly independent) to Eq.~\eqref{eq:linearsystem}, and let $c_{jk}$ be a constant $M\times M$ matrix. Then function 
\begin{equation}
\label{eq:Grammian}
u(x, y, t) = 2\,\frac{\partial^2 \ln{\tau}}{\partial x^2}, \quad
\tau(x, y, t) = \det\left[c_{jk}+\int\limits_{-\infty}^x \psi_j(x',y,t)\,\opsi_k(x',y,t)\,dx'\right].
\end{equation}
is a solution of the KP1 equation~\eqref{DimLesKPeq}. The solution is nonsingular if the determinant is everywhere positive, and we will verify this condition each time we apply this formula.

To construct solutions of Eq.~\eqref{eq:linearsystem}, let us denote 
\begin{equation}
\phi(x,y,t,\la)=\la x+i\la^2 y-4\la^3t,
\end{equation}
then for any value of $\la$ function $\psi(x,y,t) = e^{\phi(x,y,t,\la)}$ satisfies Eq.~\eqref{eq:linearsystem}. More generally, let $p_s(x,y,t,\la)$ denote the homogeneous polynomial of degree $s$ in $x$, $y$, and $t$ defined by the formula:
\begin{equation}
p_s(x,y,t,\la) = e^{-\phi(x,y,t,\la)}\frac{\partial^{s}}{\partial \lambda^s}
e^{\phi(x,y,t,\la)},
\end{equation}
where $p_0=1,\quad p_1=x+2i\la y-12\la^2t,\quad p_2 = p_1^2 + 2iy - 24\la t,\;\ldots$\\
It is easy to see that any function of the form $p_s(x,y,t,\la)e^{\phi(x,y,t,\la)}$ satisfies Eq.~\eqref{eq:linearsystem}. Using Eq.~\eqref{eq:Grammian}, we construct a wide family of solutions of the KP1 equation by choosing functions $\psi_j$ as linear combinations of the functions $p_s(x,y,t,\la)e^{\phi(x,y,t,\la)}$ for various values of $s$ and $\lambda$.

To simplify the exposition, we assume that all $\la_j=a_j>0$ are real-valued. Let $s_1,\ldots,s_M$ be non-negative integers, and let $b_1,\ldots,b_M$ be real constants. We set
\begin{equation}
    \psi_j(x,y,t) = b_j\,p_{s_j}(x,y,t,a_j)e^{\phi(x,y,t,a_j)},\quad j=1,\ldots,M.
\end{equation}
In this case, one can readily calculate:
\begin{equation}
\int\limits_{-\infty}^x \psi_j(x',y,t)\,\opsi_k(x',y,t)\,dx'=\frac{b_jb_k}{a_j+a_k}p_{s_js_k}(x,y,t,a_j,a_k)\,e^{\phi(x,y,t,a_j)+\ophi(x,y,t,a_k)}
\end{equation}
where $p_{jk}(x,y,t,a,a')$ is the non-homogeneous polynomial of degree $j+k$ given by the equation:
\begin{equation}
p_{jk}(x,y,t,a,a')=\sum_{l=0}^{j+k}\left(\frac{-1}{a+a'}\right)^l\frac{\pa^l}{\pa x^l}[p_j(x,y,t,a)\,\overline{p}_k(x,y,t,a')].
\end{equation}
For future use, we present the first few $p_{jk}$:
\begin{equation}
p_{00}=1,\quad p_{01}=x-12a^2t-\frac{1}{a+a'}-2iay,\quad
p_{10}=x-12a'^2t-\frac{1}{a+a'}+2iay,
\label{eq:p00}
\end{equation}
\begin{equation}
\left.p_{11}\right|_{a'=a}=\left(x-12a^2t-\frac{1}{2a}\right)^2+4a^2y^2+\frac{1}{4a^2}.
\label{eq:p11}
\end{equation}
Plugging this into Eq.~\eqref{eq:Grammian} and factoring out a common exponential (which disappears after application of the second logarithmic derivative), we obtain the solution in the form:
\begin{equation}
u(x,y,t) = 2\,\frac{\partial^2}{\partial x^2}
\ln \det\left[c_{jk}e^{-(\phi(x,y,t,a_j)+\ophi(x,y,t,a_k))}+\frac{b_jb_k}{2a}p_{s_js_k}(x,y,t,a_j,a_k)\right].
\label{eq:umain}
\end{equation}

We now describe several families of solution of the KP1 equation~\eqref{DimLesKPeq} given by the above formula. These solutions describe elementary interaction processes between travelling waves known as \emph{line solitons}, and localized disturbances known as \emph{lumps}. 

\begin{enumerate}
    \item \emph{Line soliton.} 
    
The standard line soliton solution of the KP1 equation is obtained by setting $M=1$, $a_1=a$, $c_{11}=1$, and $s_1=0$. In this case, the solution is:
\begin{equation}
u(x,y,t) = 2\,\frac{\partial^2}{\partial x^2}
\ln{\left(e^{-2F(x,t,a)}+C_0\right)} = 2a^2\sech^2{\left(ax - 4a^3t + \frac{1}{2}\ln{C_0}\right)},
\quad C_0=\frac{b_1^2}{2a},
\end{equation}
where
\begin{equation}
F(x,t,a)=ax-4a^3t.
\end{equation}
The terms $e^{-2F}$ and $C_0$ in the logarithm are equal on the vertical line $x=4a^2t-\frac{1}{2}\ln C_0$. The solution is supported on a narrow strip centered on this line (see left panel in Fig. \ref{f02} below); away from the line, one of the two terms in the logarithm is dominant and $u(x,y,t)$ is exponentially small. Hence the solution represents a solitary wave travelling to the right with speed $V= 4a^2$. 

\item \emph{The one-lump solution.} 

If we set $c_{jk}=0$, then the $\tau$-function in Eq.~\eqref{eq:umain} is a polynomial. The solution $u(x,y,t)$ is a rational function and consists of a collection of localized lumps, which may be bound or undergo anomalous scattering. To obtain the simplest one-lump solution, we set $M=1$, $a_1=a$, and $s_1=1$, so that
\begin{equation}
u(x,y,t) = 2\,\frac{\partial^2}{\partial x^2}\ln{\tau_l(x,y,t,a)},
\label{eq:ulump}
\end{equation}
where the $\tau$-function $\tau_l$ of the lump is given by Eq.~\eqref{eq:p11}:
\begin{equation}
\tau_l(x,y,t,a)=p_{11}(x,y,t,a,a)=\left(x-12a^2t-\frac{1}{2a}\right)^2+4a^2y^2+\frac{1}{4a^2}.
\label{eq:taulump}
\end{equation}
The solution is essentially the same as in Eq. (\ref{Lump}); it is centered at the point $(x,y)=(12a^2t+1/2a,0)$ and represents a lump travelling to the right with the speed $V = 12a^2$. The solution has a local maximum at the center and decays algebraically as $(x^2+y^2)^{-1}$ at infinity (see Fig. \ref{f01}a).

\item \emph{Line soliton absorbing or emitting a lump.} 

Let us consider the following set of parameters: $M=2$, $a_1=a_2=a$, $s_1=0$, and $s_2=1$ and assume that $c_{jk}$ is the rank one matrix:
\begin{equation}
c_{jk}=\left(\begin{array}{cc}
1 & 0 \\ 0 & 0\end{array}\right).
\label{eq:crankone}
\end{equation}
In this case, the solution reads (after factoring out a constant):
\begin{equation}
u(x,y,t) =
2\,\frac{\partial^2}{\partial x^2} \ln\left[\tau_l(x,y,t,a)+C_1e^{2F(x,t,a)}\right],\quad C_1=\frac{b_1^2}{8a^3}
\label{eq:solitonabsorbinglump}
\end{equation}
In contrast to the two previous cases, this solution is non-stationary. For a fixed moment of time $t$, consider the curve $2F=\ln \tau_l-\ln C_1$ in the $(x,y)$-plane along which the two terms in the logarithm are equal. To the right of this curve, the dominant term is $e^{2F}$, and the solution $u$ is exponentially small. To the left, the dominant term $\tau_l$ (which is the $\tau$-function~\eqref{eq:taulump} of the lump) produces a lump solution, moving along the $x$-axis with the speed $V_l = 12a^2$. Along the curve, there is a line soliton (deformed near $y=0$ by the $\tau_l$ term), moving to the right with the speed $V_s=4a^2$. Since the lump moves with the speed $V_l = 3V_s$, it eventually merges with the line soliton and is absorbed by it. 
One can also construct a line soliton emitting a lump; this situation occurs if we set $M=2$, $a_1=a_2=a$, $s_1=s_2=1$, and $c_{jk}=\delta_{jk}$, where $\delta_{jk}$ is the Kronecker symbol. Indeed, in this case, we factor out the exponential function $e^{-2F}$, which disappears after differentiation, and obtain:
\begin{equation}
u(x,y,t) =
2\,\frac{\partial^2}{\partial x^2} \ln\left[e^{-2F(x,t,a)}+\frac{b_1^2+b_2^2}{2a}\tau_l(x,y,t,a)\right].
\label{eq:solitonemittinglump}
\end{equation}
The solution again consists of a line soliton moving with the speed $V_s=4a^2$ along the curve where the two terms are equal, and a lump moving with speed $V_l=12a^2$. However, unlike the previous case, the term $\tau_l$ that produces a lump is now dominant to the right of the line soliton. Hence, the line soliton radiates a lump in the process of evolution. Solution (\ref{eq:solitonemittinglump}) is illustrated by the contour lines in Fig. \ref{f02} at two time instances with the following parameters: $a = 1$, $b_1 = 0$ and $b_2 = 10^6$. The line soliton with the bent front shown in Fig. \ref{f02}b asymptotically becomes straight like in Fig. \ref{f02}a. 
\begin{figure}[h!]
\centering
\includegraphics[width=1.0\textwidth]{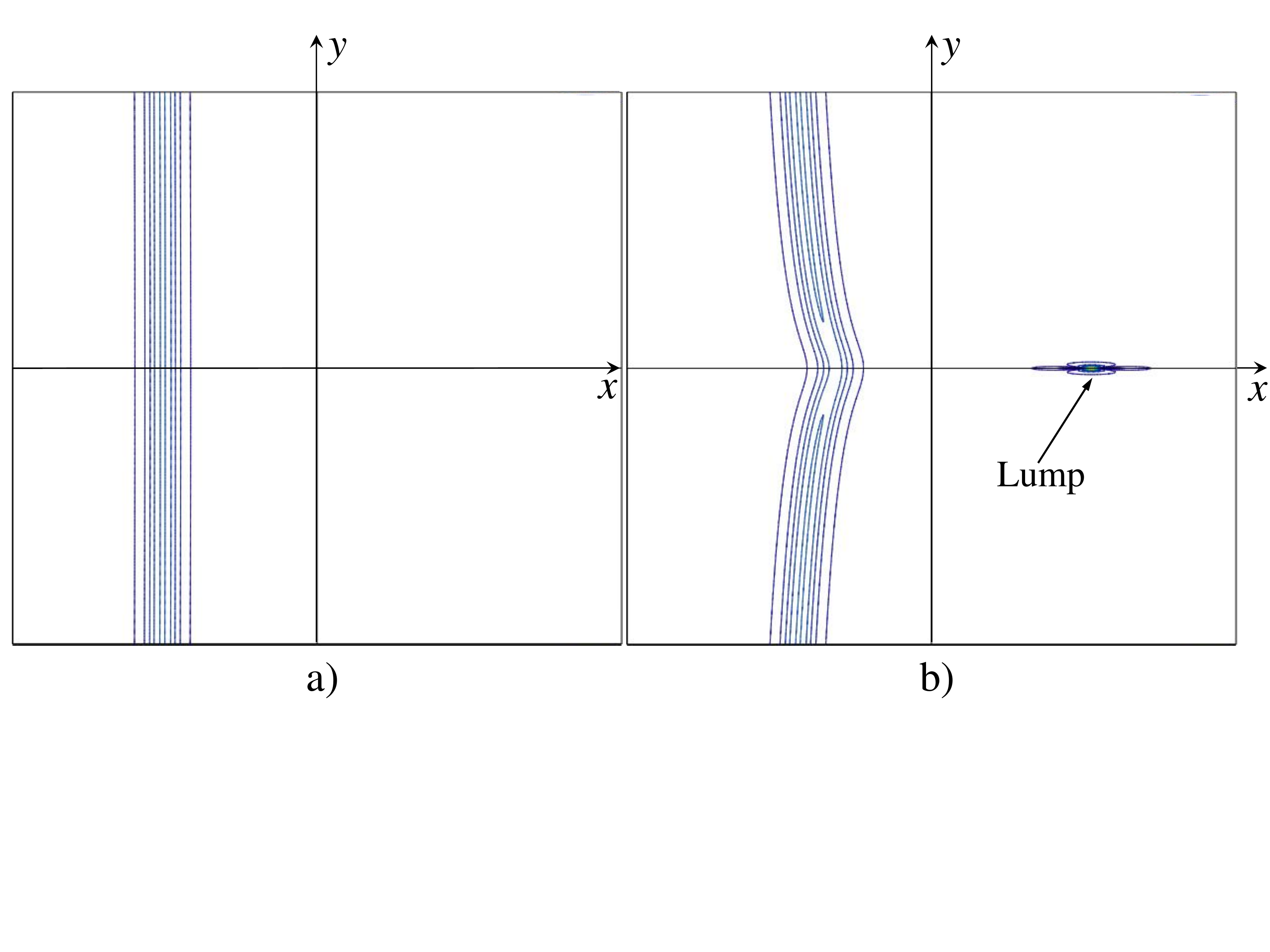}
\vspace{-4.5cm}
\caption{Slightly perturbed line soliton at $t = -100$ (panel a) radiating a lump (panel b at $t = 0$). The vertical scale in both panels is 5 times greater than the horizontal scale. Panel (a) $-430 \le x \le -390$; panel (b) $-30 \le x \le 10$. In both panels $-100 \le y \le 100$.} %
\label{f02}
\end{figure}

\item \emph{Lump passing through a line soliton.} 

As has been demonstrated above, a KP lump moving with the speed $V_l$ is in resonance with a line soliton moving with the speed $V_s$ if $V_l/V_s=3$. If $V_l/V_s\neq 3$, then the lump can pass through the line soliton or may create together with it a stationary pattern when $V_l/V_s=1$.

To construct such a solution, let us set $M=2$, $a_2\neq a_1$, $s_0=1$, $s_1=1$, and assume that $c_{jk}$ is the rank one matrix~\eqref{eq:crankone}. A direct calculation shows that
the corresponding solution is:
\begin{equation}
\label{Eq.25}
u(x,y,t) = 2\,\frac{\partial^2}{\partial x^2} \ln \left[
\tau_l(x,y,t,a_2)+C_2\tau_l(x-x_0,y,t,a_2)e^{2F(x,t,a_1)}\right],
\end{equation}
where
\begin{equation}
C_2=\frac{(a_1-a_2)^2b_1^2}{a_1(a_1+a_2)^2},\quad x_0=\frac{2a_1}{a_1^2-a_2^2}.
\end{equation}

There is a line soliton along the vertical line $2F=-\ln C_2$ moving with the speed $V_s=4a_1^2$. On either side of the soliton, there is a one-lump solution moving with the same speed $V_l=12a_2^2$ and having a relative phase shift $x_0$ between them.

If $a_2<a_1<a_2\sqrt{3}$, then the solution consists of a lump overtaking a line soliton from the left, and reappearing to the right of the line soliton with the phase shift $x_0$. If $a_1>a_2\sqrt{3}$, then it is instead the line soliton that overtakes the lump, which reappears behind it. In either case, there is a relatively short time interval during which both lumps are visible. This situation is demonstrated in Fig. \ref{f06}.
\begin{figure}[h!]
\centering
\includegraphics[width=1.0\textwidth]{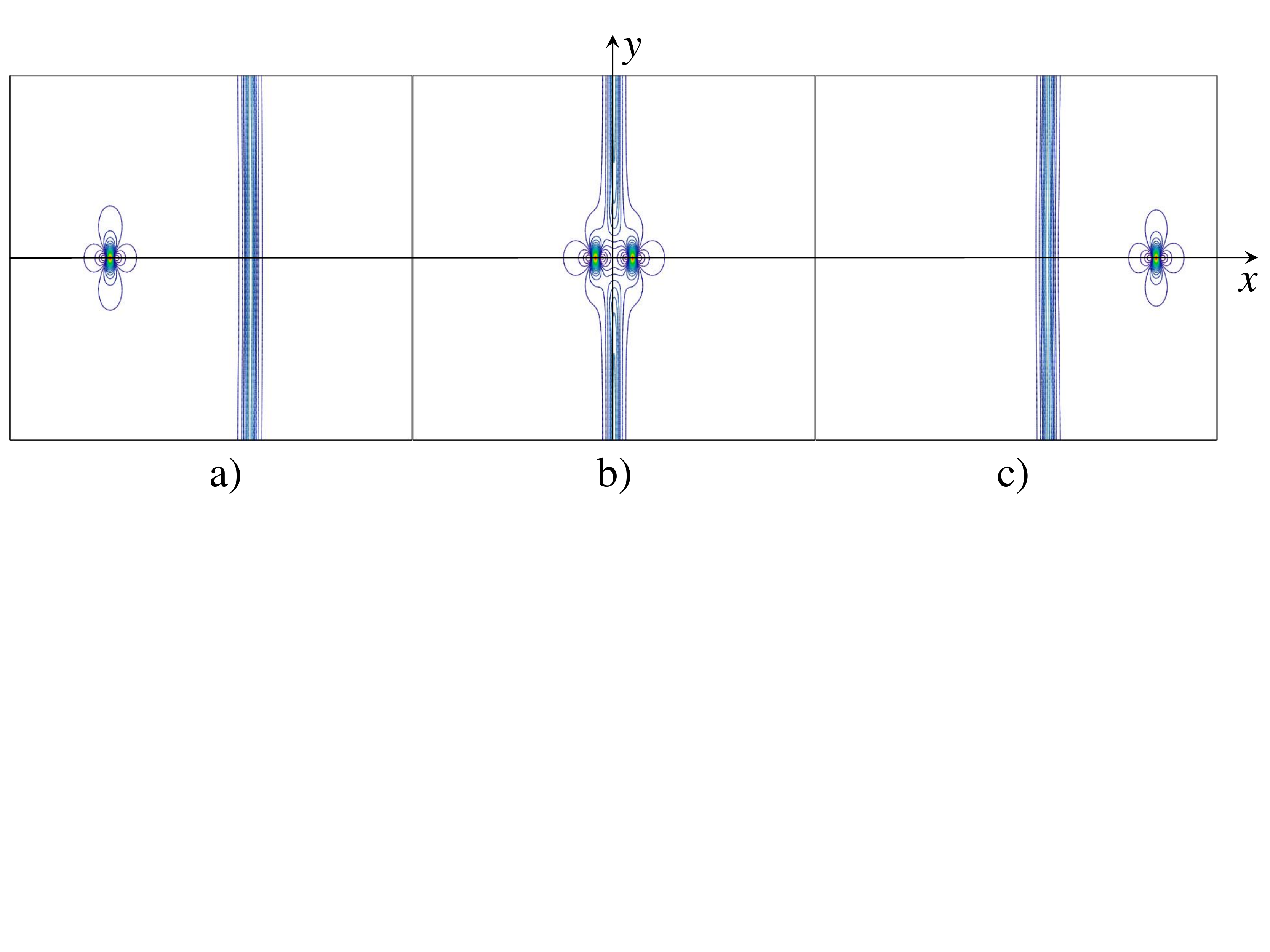}
\vspace{-7.cm}
\caption{Line soliton overtaking by a lump as per Eq. (\ref{Eq.25}). Frame (a), $t = -200$, $-500 \le x \le -300$; frame (b), $t = -85$, $-175 \le x \le 25$; frame (c), $t = 0$, $-150 \le x \le 50$; in all frames $-100 \le y \le 100$.} %
\label{f06}
\end{figure}

In the boundary case $a_1=a_2\sqrt{3}$, the two lumps are traveling with the same speed as the line soliton, and the solution is stationary. The stationary solution is shown in Fig. \ref{f05} for the following set of parameters: 
$$
a_1 = \frac{1}{2}, \quad a_2 = \frac{1}{\sqrt{12}}, \quad b_1 = \frac{\left(\sqrt{3} + 1\right)^2}{2}\exp{\left[-\frac{\sqrt{3}}{2}\left(\sqrt{3} + 1\right)\right] \approx 0.35}, \quad b_2 = \sqrt[4]{3}.
$$
In this case, the $\tau$-function is:
\begin{equation}
\label{Bonded-F}
\tau(\xi,y) = 3(\xi+3)^2 + y^2 + 9 + e^\xi\left[3(\xi-3)^2 + y^2 + 9\right],
\end{equation}
where $\xi = x - Vt$, and $V = 1$. This and some other stationary solutions were  derived in \cite{Abramyan-1985}. In particular, solutions representing a chain of lumps intersecting with a plane soliton at an angle were found in that paper; they will be described in Section \ref{Sect-3}.
\begin{figure}[h!]
\centering
\includegraphics[width=1.0\textwidth]{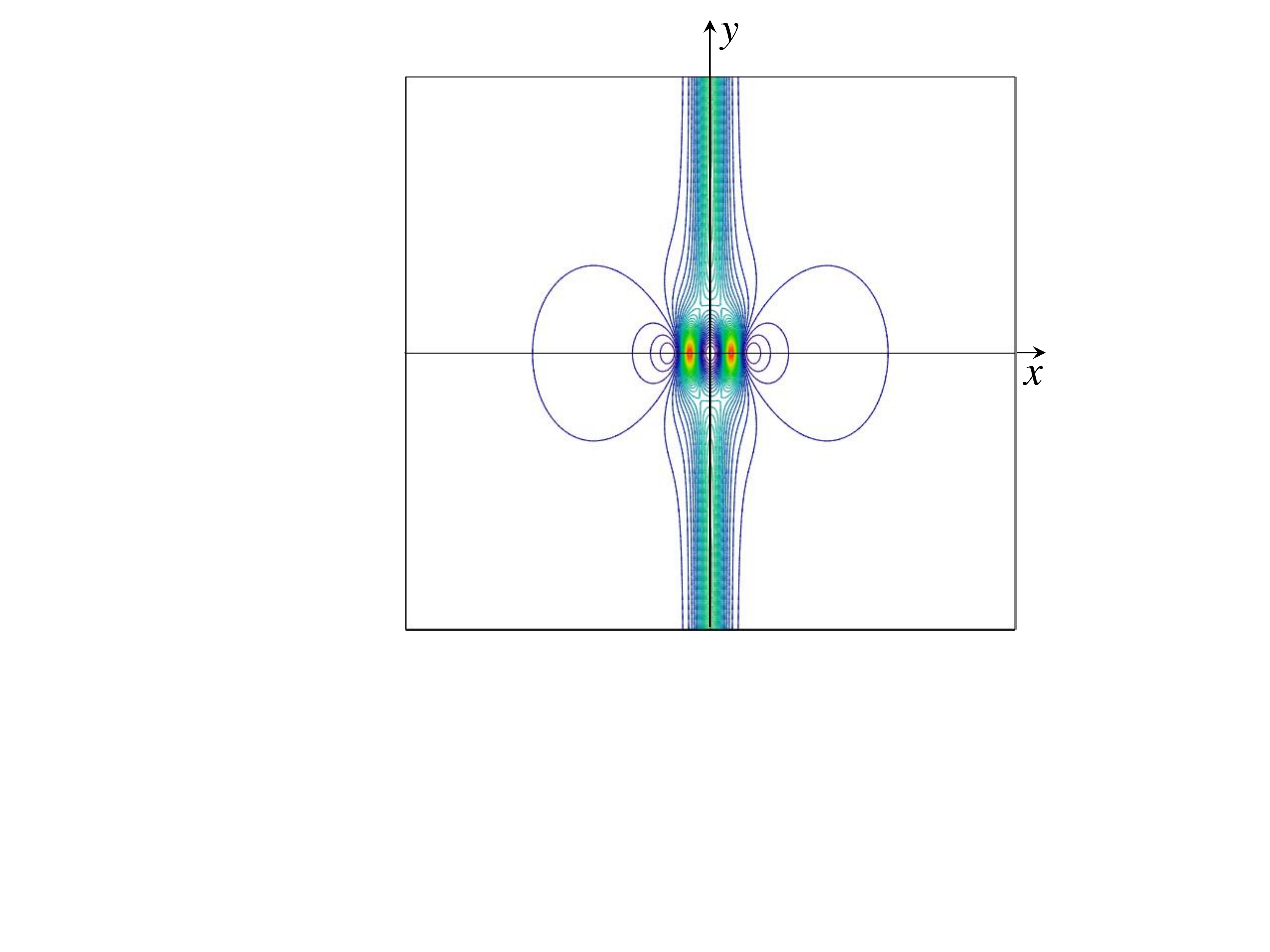}
\vspace{-5.0cm}
\caption{Two lumps of equal amplitudes bonded with a line soliton. The vertical and horizontal sizes of the frame are equal: $-50 \le x \le 50$; $-50 \le y \le 50$.} %
\label{f05}
\end{figure}

As we approach the resonance case $a_1\to a_2$, the phase shift $x_0$ diverges, the line soliton and the right lump disappear at plus infinity; the only the left lump remains. For $a_1<a_2$, the solution again consists of a lump overtaking a line soliton. \\ 

\item \emph{Two weakly bound line solitons exchanging a lump.} 

Consider now solution~\eqref{eq:umain} with $M=2$, $a_1=a_2=a$, $s_1=0$, $s_2=1$, $c_{jk}=\delta_{jk}$. In this case, a direct calculation shows that:
\begin{equation}
u(x,y,t) =
2\,\frac{\partial^2}{\partial x^2} \ln\left[
e^{-2F(x,t,a)}+Q(x,y,t)+C_3e^{2F(x,t,a)}\right],
\label{eq:utwosolitons}
\end{equation}
where $Q$ and $C_3$ are expressed in terms of the one-lump tau function $\tau_l$ (see~\eqref{eq:taulump}) as follows:
\begin{equation}
Q(x,y,t) = \frac{b_2^2}{2a}\tau_l(x,y,t,a)+\frac{b_1^2}{2a},\quad
C_3=\frac{b_1^2b_2^2}{16a^4}.
\end{equation}

The structure of solution~\eqref{eq:utwosolitons} is determined by the relative values of the three terms under the logarithm. For a fixed moment of time $t$, the $(x,y)$-plane is divided into three regions; in each of of them one of the three terms in the logarithm in Eq.~\eqref{eq:utwosolitons} is dominant:
\begin{equation*}
\Delta_1(t)=\{(x,y):e^{-2F}\geq \max(Q,C_3e^{2F})\},\quad
\Delta_2(t)=\{(x,y):Q\geq \max(e^{-2F},C_3e^{2F})\},
\end{equation*}
\begin{equation}
\Delta_3(t)=\{(x,y):C_3e^{2F}\geq \max(e^{-2F},Q)\}.
\end{equation}

To determine the shape of the regions $\Delta_i(t)$, consider the three functions $e^{-2F}$, $Q$, and $C_3e^{2F}$ on the $x$-axis for fixed values of $y$ and $z$. It is clear that $e^{-2F}$ is dominant when $x\to -\infty$, whereas $C_3e^{2F}$ is dominant when $x\to +\infty$. In the meantime, there is always an intermediate domain where $Q$ is the dominant term. Indeed, the minimum value of the function $\max(e^{-2F},C_3e^{2F})$ is $\sqrt{C_3}$, whereas
\begin{equation}
Q(x,y,t)\geq \frac{1}{2}\left(\frac{b_2^2}{4a^3}+\frac{b_1^2}{a}\right)\geq
\sqrt{\frac{b_2^2}{4a^3}\cdot\frac{b_1^2}{a}}=2\sqrt{C_3}.
\end{equation}
The length of this intermediate interval on the $x$-axis is smallest when $y=0$ and increases logarithmically when $|y| \to \infty$. It follows that the region $\Delta_2$ is a vertical strip separating $\Delta_1$ on the left and $\Delta_3$ on the right. 

Thus, solution~\eqref{eq:utwosolitons} is as follows. In the interior of the regions $\Delta_1$ and $\Delta_3$, where one of the exponential functions, either $e^{-2F}$ or $C_3e^{2F}$, is the dominant in Eq.~\eqref{eq:umain}, the solution is exponentially small. In the domain $\Delta_2$, the solution is approximately equal to $u \approx 2\,\partial^2\ln{Q}/\partial x^2$. In the limiting case $b_1/b_2\to 0$ this is a lump, travelling right with velocity $V_l=12a^2$, while for $b_1>b_2$ the lump is flattened and becomes invisible when $b_1/b_2\to +\infty$. There are two line solitons along the curves $\Delta_1\cap \Delta_2$ and $\Delta_2\cap \Delta_3$, both travelling to the right with the same velocity $V_s=4a^2$. The solitons are deformed and are closest at $y=0$, and the distance between them increases logarithmically as $|y|\to +\infty$. Since $V_l=3V_s$ as before, the lump is emitted by the left line soliton and absorbed by the right line soliton. We note that solution~\eqref{eq:solitonabsorbinglump} can be obtained in the limit $b_2\to +\infty$, while solution~\eqref{eq:solitonemittinglump} corresponds to sending $b_1\to 0$; in both limits one of the two line solitons disappears to infinity.

The solution is shown in Fig. \ref{f03} for $a = 1$, $b_1 = 10^2$, and $b_2 = 10^6$ and in Fig. \ref{f04} for $a = 1$, $b_1 = b_2 = 1$. In the former case, a lump is clearly seen between two line solitons (see Fig. \ref{f03}b), whereas in the latter case, the lump is flattened and cannot be distinguished from the interacting line solitons. 
\begin{figure}[h!]
\centering
\includegraphics[width=0.88\textwidth]{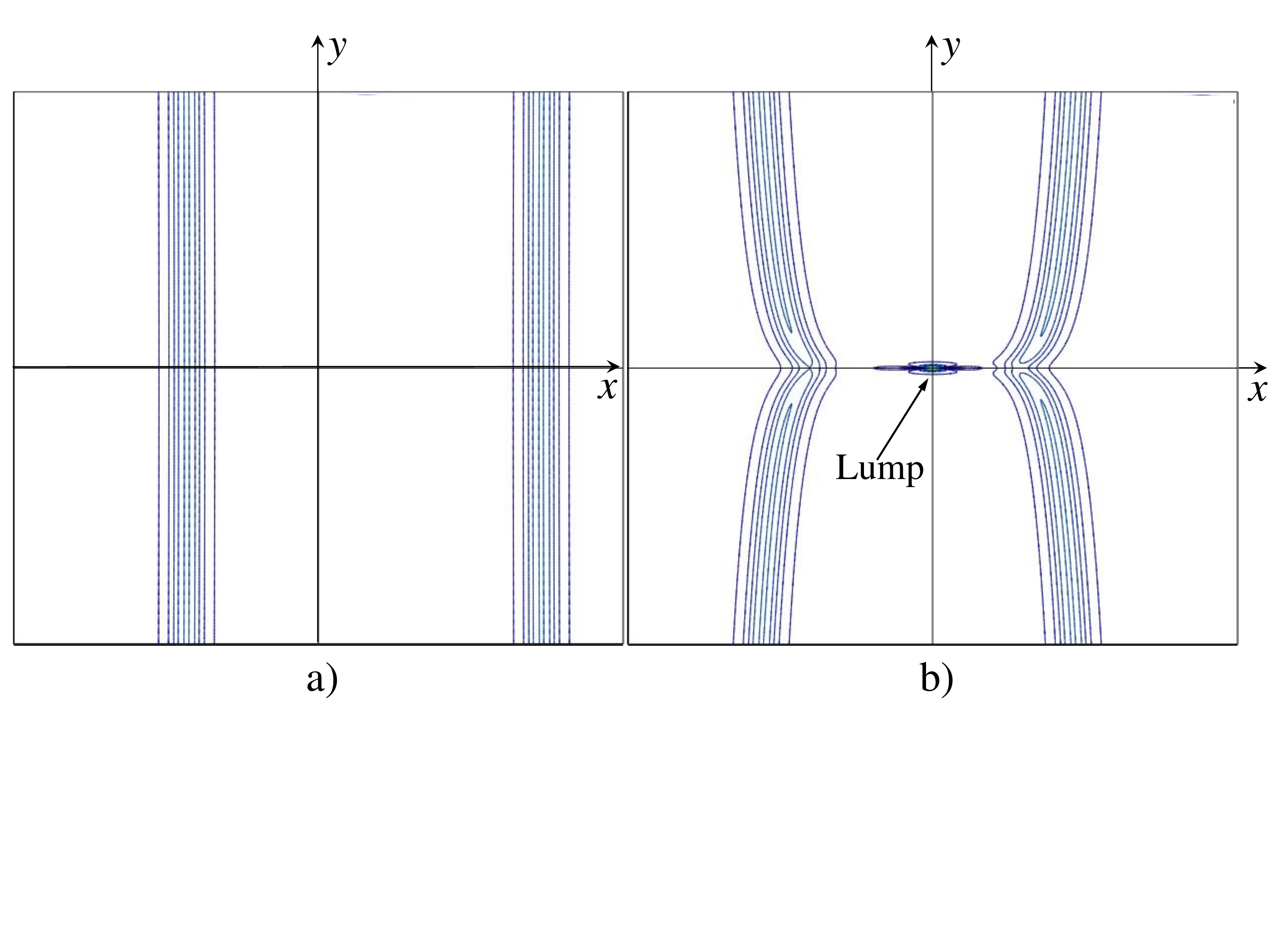}
\vspace{-3.5cm}
\caption{Two quasi-parallel solitons of equal amplitudes at $t = -100$ (panel a). In the process of evolution, the left soliton radiates a lump shown in the panel (b) at $t = -1$ which is eventually absorbed by the right soliton. The vertical size in both panels is 5 times longer than the horizontal size. Panel (a): $-431.5 \le x \le -391.5$; panel (b): $-31.5 \le x \le 8.5$. In both panels $-100 \le y \le 100$.} %
\label{f03}
\end{figure}
\begin{figure}[h!]
\centering
\includegraphics[width=0.88\textwidth]{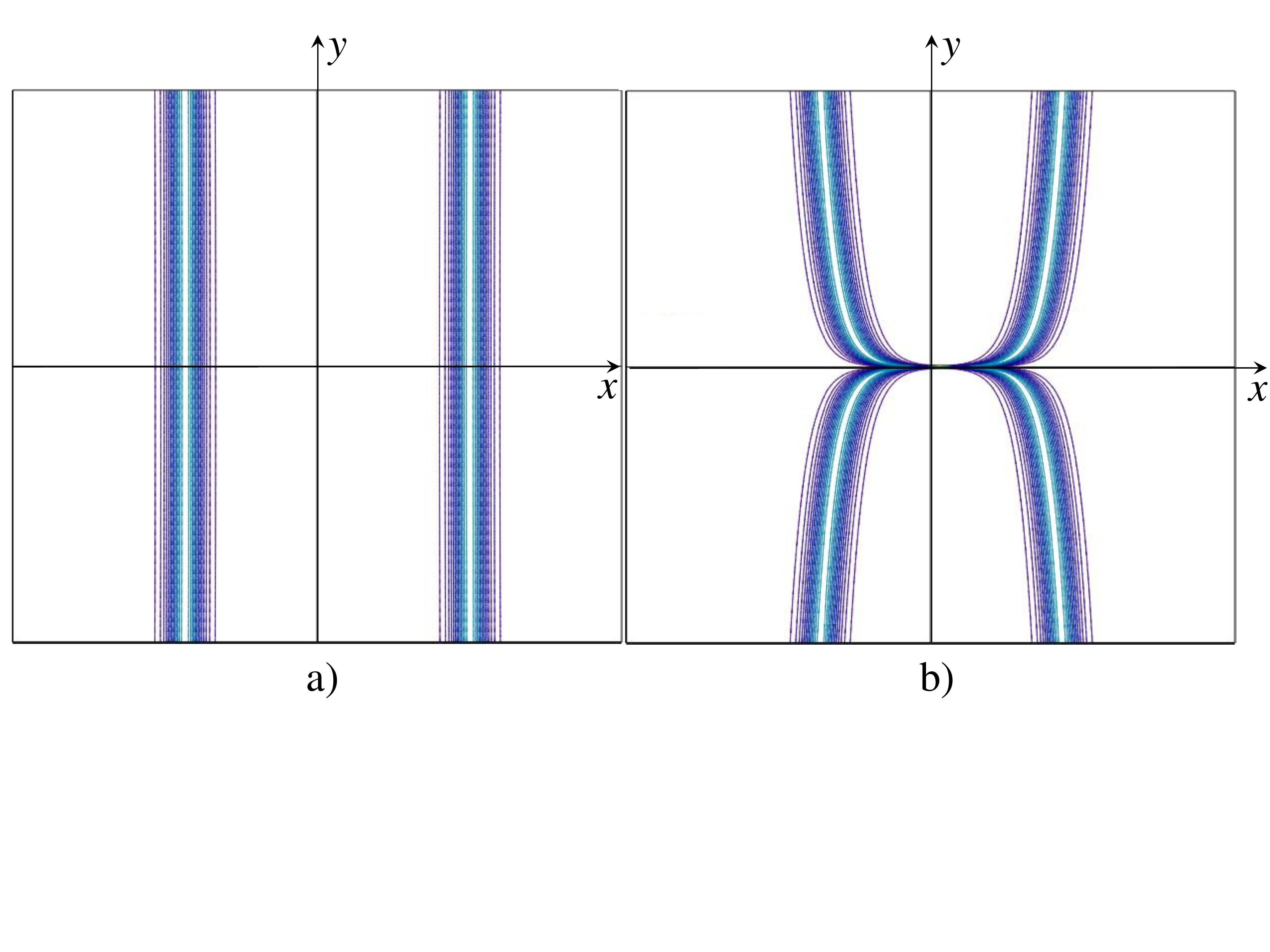}
\vspace{-3.5cm}
\caption{Two quasi-parallel solitons of equal amplitudes at $t = -1000$ (panel a) and at $t = 0$ (panel b). A flat-head lump is obscured between the solitons in the field of their tails. The vertical size in both panels is 50 times longer than the horizontal size. Panel (a): $-4020 \le x \le -3980$; panel (b): $-20 \le x \le 20$. In both panels $-1000 \le y \le 1000$.} %
\label{f04}
\end{figure}

We now interpret our solution. Two parallel KP1 line solitons of equal amplitude separated by a large distance can be described by the KdV equation in the ideal case when their fronts are not perturbed along the $y$-direction. In such a case, they will experience an exchange-type interaction: a portion of the energy from the rear soliton will be transmitted to the front soliton. Since the front soliton now has a larger amplitude, the distance between the solitons will diverge linearly in time, proportionally to the amplitude difference.

The solution that we construct consists of two parallel line solitons of equal amplitude that are infinitesimally perturbed at $t\to -\infty$. The one-dimensional KdV equation is insufficient in this case, and the two-dimensional KP1 equation must be used. The distance between the solitons decreases logarithmically with time until the solitons exchange a lump, whereupon the solitons diverge logarithmically with time. At $t \to +\infty$, we have again two parallel line solitons of equal amplitudes.

\end{enumerate}

\section{INTERACTION OF A LINE SOLITON AND LUMP CHAINS}
\label{Sect-3}

In Section \ref{Sect-2}, we presented solutions of KP1 consisting of line solitons interacting with individual lumps. The KP1 equation also has solutions describing infinite sequences of lumps, known as \emph{lump chains} \cite{Zaitsev-1983, Gdanov-1984, Lester-2021}. Lump chains can interact with line solitons in a manner similar to individual lumps. Below we present several examples of such interactions in terms of the Grammians and corresponding $\tau$-functions.

\begin{enumerate}
\item \emph{Lump chain.} 

To construct an isolated lump chain, we follow the method of~\cite{Lester-2021} and consider the Grammian $\tau$-function~\eqref{eq:Grammian} with $M=1$ and $c_{11}=0$. Let us choose two real-valued spectral parameters $a_1<a_2$ and real-valued phases $\rho_1$, $\rho_2$, and choose the $\psi$-function in the form:
\begin{equation}
\psi(x,y,t)=\psi_1(x,y,t)=\sqrt{2a_1}e^{\phi(x,y,t,a_1)+\rho_1}+\sqrt{2a_2}e^{\phi(x,y,t,a_2)+\rho_2}. 
\end{equation}
Then, we can calculate the corresponding $\tau$-function as:
\begin{equation}
\tau_c(x,y,t) = \int_{-\infty}^x|\psi(x',y,t)|^2dx =e^{2F_1}+e^{2F_2}+C_4e^{F_1+F_2}\cos{\left[\left(a_1^2-a_2^2\right)y\right]}, \label{eq:tauc}
\end{equation}
where
\begin{equation}
F_1(x,t)=a_1x-4a_1^3t+\rho_1,\quad F_2(x,t)=a_2x-4a_2^3t+\rho_2,\quad
C_4=4\frac{\sqrt{a_1a_2}}{a_1+a_2}.
\end{equation}
The two exponentials $e^{2F_1}$ and $e^{2F_2}$ are equal along the vertical line:
\begin{equation}
x=V_ct-\rho_{12},\quad V_c=4(a_1^2+a_1a_2+a_2^2),\quad \rho_{12}=\frac{\rho_1-\rho_2}{a_1-a_2}.
\label{eq:F1=F2}
\end{equation}
Away from this line, one of the two exponentials is dominant and the corresponding solution $u(x,y,t)$ is exponentially small. Meanwhile, it can be shown that the term containing the cosine function is never dominant. Hence, the solution is concentrated in a narrow strip along this line, and consists of an evenly spaced sequence of lumps. The lumps propagate to the right with the speed $V_c$. By an appropriate choice of the spectral parameters, this solution can be presented in two forms \cite{Abramyan-1985}. One of them describes a line of lumps periodic in the $y$-direction and moving along the $x$-direction:
\begin{equation}
\label{Y-chain}
\tau = \cosh{\left(k\xi\sqrt{V_c}\right)} - \sqrt{\frac{1 - 4k^2}{1 - k^2}}\cos{\left(y\frac{kV_c}{\sqrt{3}}\sqrt{1 - k^2}\right)},
\end{equation}
where $\xi = x - V_ct$ and $|k| < 1/2$. Another solution can be obtained from the previous one by replacing $k = i\kappa$; it describes a line of lumps periodic in the direction of motion, the $x$-direction:
\begin{equation}
\label{X-chain}
\tau = \cos{\left(\kappa\xi\sqrt{V_c}\right)} + \sqrt{\frac{1 + 4\kappa^2}{1 + \kappa^2}}\cosh{\left(y\frac{\kappa V_c}{\sqrt{3}}\sqrt{1 + \kappa^2}\right)}.
\end{equation}
Both these solutions reduce to the solution for a lump when $k \to 0$ or $\kappa \to 0$; then the degenerate solution becomes: 
\begin{equation}
\label{Tau-lump}
\tau = 3V\xi^2 + V^2y^2 + 9.
\end{equation}

\item \emph{Line soliton radiating a lump chain.} 

Let us choose again two real-valued spectral parameters $a_1<a_2$ and real-valued phases $\rho_1$, $\rho_2$, but now we modify the $\tau$-function~\eqref{eq:tauc} by setting $c_{11}=1$:
\begin{equation}
\label{Eq.39}
u(x,y,t)=
2\,\frac{\partial^2}{\partial x^2}\ln \left(1+e^{2F_1}+e^{2F_2}+C_4e^{F_1+F_2}\cos{\left[\left(a_1^2-a_2^2\right)y\right]}\right).
\end{equation}
The new $\tau$-function $\tau(x,y,t)=1+\tau_c(x,y,t)$ has three terms that may be dominant: $1$, $e^{2F_1}$, and $e^{2F_2}$. The relative values of these terms depend on the relationship between $x$ and $t$:
\begin{center}
\begin{tabular}{c c c c c c}
$e^{2F_2(x,t)}>e^{2F_1(x,t)}$ & if & $x>V_ct-\rho_{12}$, &
$e^{2F_2(x,t)}<e^{2F_1(x,t)}$ & if & $x<V_ct-\rho_{12}$,\\
$e^{2F_1(x,t)}>1$ & if & $x>4a_1^2t-\rho_1/a_1$, &
$e^{2F_1(x,t)}<1$ & if & $x>4a_1^2t-\rho_1/a_1$,\\
$e^{2F_2(x,t)}>1$ & if & $x>4a_2^2t-\rho_2/a_2$, &
$e^{2F_2(x,t)}<1$ & if & $x>4a_2^2t-\rho_2/a_2$.
\end{tabular}
\end{center}

Since $4a_1^2<4a_2^2<V_c \equiv 4(a_1^2+a_1a_2+a_2^2)$, for sufficiently large positive $t$ the dominant terms in the $\tau$-function are as follows (going left to right):
\begin{equation}
1\mbox{ if }x<4a_1^2t-\rho_1/a_1,\quad
e^{2F_1}\mbox{ if } 4a_1^2t-\rho_1/a_1<x<V_ct-\rho_{12}, \quad e^{2F_2}\mbox{ if }x>V_ct-\rho_{12}. 
\end{equation}
Along the vertical line $x=4a_1^2t-\rho_1/a_1$ we have $1=e^{2F_1}\gg e^{2F_2}$; this corresponds to a line soliton moving with the speed $V_{l1} = 4a_1^2$. Similarly, along the vertical line $x=4a_1^2t-\rho_1/a_1$, we have $e^{2F_1}=e^{2F_2}\gg 1$, so that $\tau(x,y,t)\simeq \tau_c(x,y,t)$, which corresponds to a lump chain moving with the speed $V_c$. Away from these two lines, the solution is exponentially small.

For sufficiently large negative $t$ the dominant terms in the $\tau$-function are:
\begin{equation}
1\mbox{ if }x<4a_2^2t-\rho_2/a_2,\quad
e^{2F_2}\mbox{ if }x>4a_2^2t-\rho_2/a_2.
\end{equation}
This corresponds to a line soliton moving with the speed $V_{l2} = 4a_2^2$ along the vertical line $x>4a_2^2t-\rho_2/a_2$ where $1=e^{2F_2}\gg e^{2F_1}$. The solution is exponentially small away from this line.

Hence, the solution has the following structure. For $t<0$ there is a line soliton traveling with speed $V_{l2} = 4a_2^2$. At a certain moment of time, the soliton radiates a vertical chain of lumps periodic in the $y$-direction, which propagates away from the original soliton with the speed $V_c>V_{l2}$ and another line soliton moving with the slower speed $V_{l1} = 4a_1^2$. Such solution was obtained for the first time in \cite{Pelinovsky-1993B}. We note that if the spectral parameters $\la_1$ and $\la_2$ are complex-valued, then the solution consists of a bent line soliton emitting a lump chain at an angle (see~\cite{Lester-2021}). As $\la_1$ and $\la_2$ become real-valued, the triple point at which the lump chain is emitted moves off to infinity.

\item \emph{Line soliton absorbing a lump chain.} 

There is an inverse process when a line soliton absorbs a periodic lump chain. To derive such a solution, we choose the following parameters: $a_1<a_2$, real phases $\rho_1$ and $\rho_2$, $M=2$, and assume that $c_{jk}$ is the rank one matrix with $c_{22}=1$ and all other entries $c_{jk}=0$. Then, we choose the following auxiliary functions:
\begin{equation}
\psi_1(x,y,t)=\sqrt{2a_1}e^{\phi(x,y,t,a_1)}+\sqrt{2a_2}e^{\phi(x,y,t,a_2)},\quad \psi_2=\sqrt{2a_1}e^{\phi(x,y,t,a_1)}.
\label{eq:psiforchain}
\end{equation}
A direct calculation shows that the $\tau$-function in this case is:
\begin{equation}
\label{eq:C4}
\tau=e^{2F_1}+e^{2F_2}+C_4e^{F_1+F_2}\cos{\left[\left(a_1^2-a_2^2\right)y\right]}+C_5e^{2F_1+2F_2},\quad 
C_5=\frac{(a_1-a_2)^2}{(a_1+a_2)^2}.
\end{equation}
The three terms to be compared in the $\tau$-function are $e^{2F_1}$, $e^{2F_2}$, and $C_5e^{2(F_1+F_2)}$. 
An analysis similar to the one presented above shows that for the large positive $t$ there is a single line soliton moving with the speed $V_{e2} = 4a_2^2$, while for large negative $t$ there is a line soliton moving with the speed $V_{e1} = 4a_1^2$ and a lump chain moving with the speed $V_c$. At a certain moment of time, the lump chain is absorbed by the line soliton, causing its speed to increase from $V_{e1}$ to $V_{e2}$. This process is illustrated by Fig. \ref{f08}. In frame (a) generated for $t = -30$, one can see a fragment containing two lumps in the left and a line soliton at a large distance from the lump chain on the right. In frame (b) generated for $t = 0$, one can see the lump chain approaching the line soliton which becomes noticeably modulated. In frame (c) generated for $t = 30$, there is only one line soliton which has absorbed the lump chain and now moves with the higher speed having a larger amplitude.
\begin{figure}[h!]
\centering
\includegraphics[width=1.0\textwidth]{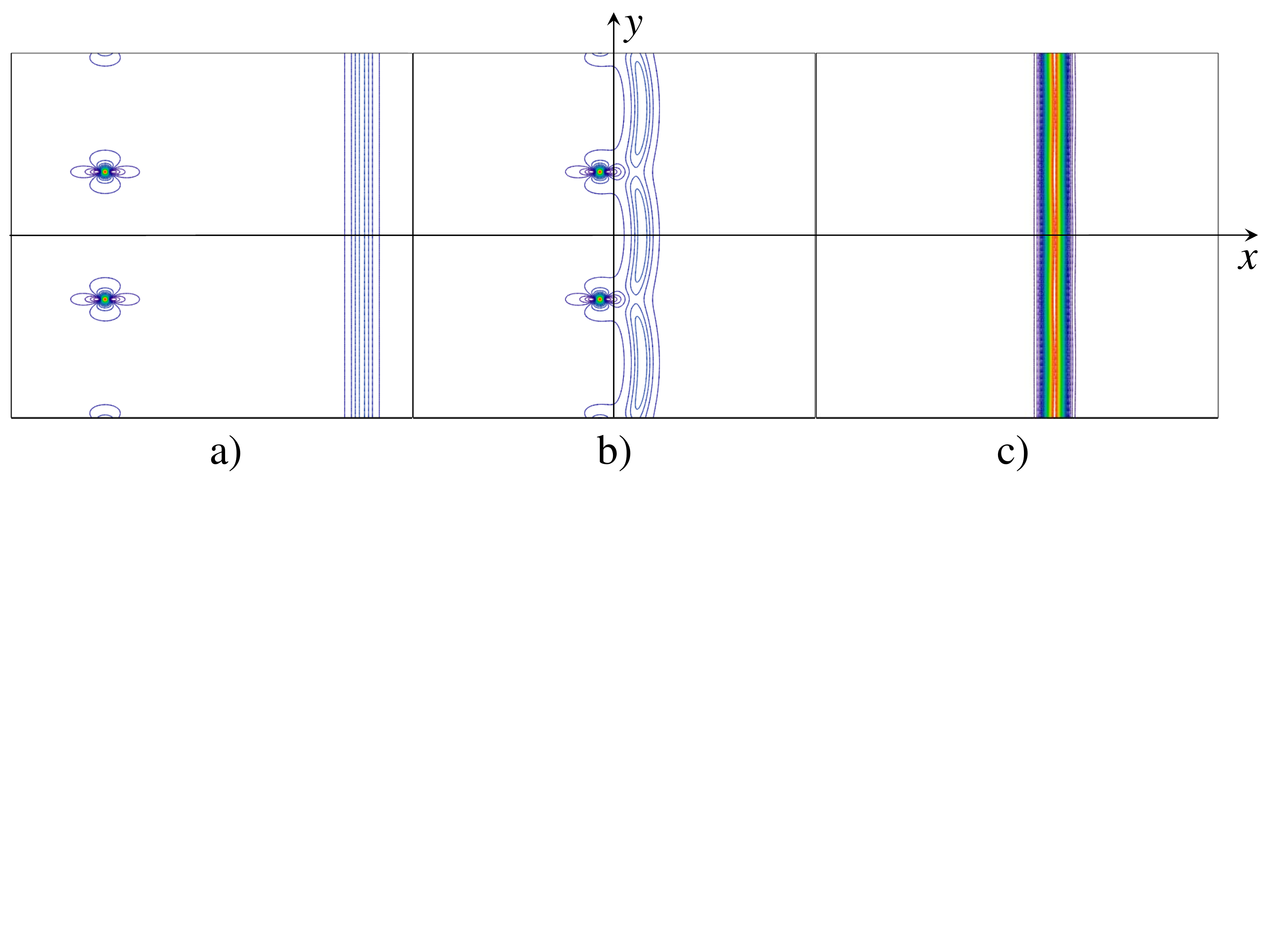}
\vspace{-7.cm}
\caption{Absorption of a periodic chain of lumps by a line soliton. Amplitudes of lumps are $A_l = 3.2$, amplitude of the initial line soliton is $A_{e1} = 2a_1^2 = 0.32$, amplitude of the resultant line soliton is $A_{e2} = 2a_2^2 = 0.5$. Frame (a), $t = -30$, $-100 \le x \le 0$; frame (b), $t = 0$, $-50 \le x \le 50$; frame (c), $t = 30$, $-25 \le x \le 75$; in all frames $-100 \le y \le 100$.} %
\label{f08}
\end{figure}

\item \emph{Line solitons exchanging a lump chain.} 

Finally, let us construct a solution consisting of two line solitons exchanging a chain of lumps in the process of their interaction. As above, we set $M=2$, choose the spectral parameters $a_1<a_2$ and real phases $\rho_1$, $\rho_2$, and $\rho_3$, and use the auxiliary functions
\begin{equation}
\psi_1(x,y,t)=\sqrt{2a_1}e^{\phi(x,y,t,a_1)+\rho_1}+\sqrt{2a_2}e^{\phi(x,y,t,a_2)+\rho_2},\quad \psi_2=\sqrt{2a_1}e^{\phi(x,y,t,a_1)+\rho_3}.
\end{equation}
We now set $c_{jk}=\delta_{jk}$. A calculation shows that the $\tau$-function in this case is:
\begin{equation}
\label{Eq.45}
\tau(x,y,t) = 1 + (1+C_6)e^{2F_1} + e^{2F_2} + C_4e^{F_1+F_2}\cos{\left[\left(a_1^2-a_2^2\right)y\right]} + C_5 C_6 e^{2F_1+2F_2},
\end{equation}
$$
\mbox{where} \quad 
C_4=4\frac{\sqrt{a_1a_2}}{a_1+a_2}, \quad C_5=\frac{(a_1-a_2)^2}{(a_1+a_2)^2}, \quad C_6=e^{2(\rho_3-\rho_1)}.
$$

The four terms in the $\tau$-function that may be dominant are $1$, $(1+C_6)e^{2F_1}$, $e^{2F_2}$, and $C_5C_6 e^{2F_1+2F_2}$, and the solution is nonvanishing near the lines along which two of these terms are equal and greater than the other two. 

To simplify exposition, we set $\rho_1= -\ln(1+C_6)/2$ and $\rho_2=0$ (this can be achieved by translating in $x$ and $t$). To find the dominant term, we solve the possible equalities (we consider only the case when $C_6$ is small, so the dominant pair cannot be $1$ and $C_5C_6 e^{2F_1+2F_2}$):
\begin{eqnarray}
1=(1+C_6)e^{2F_1}, &\mbox{ when }& x=x_1(t)=4a_1^2t, \nonumber \\
1=e^{2F_2}, &\mbox{ when }& x=x_2(t)=4a_2^2t, \nonumber \\
(1+C_6)e^{2F_1}=C_5C_6 e^{2F_1+2F_2}, &\mbox{ when }& x=x_3(t)=4a_2^2t+\frac{\ln (1+C_6^{-1})-\ln C_5}{2a_2}, \nonumber \\
e^{2F_2}=C_5C_6 e^{2F_1+2F_2}, &\mbox{ when }& x=x_4(t)=4a_1^2t+\frac{\ln (1+C_6^{-1})-\ln C_5}{2a_1}, \nonumber \\
(1+C_6)e^{2F_1}=e^{2F_2}, &\mbox{ when }& x=x_5(t)=V_ct.  \nonumber
\end{eqnarray}

Carefully analysing the relative values of the terms for different $x$, we see that there are three possibilities, depending on the time $t$:

\begin{enumerate}
    \item When $t<0$, $1$ is dominant for $x<x_2(t)$; $e^{2F_2}$ is dominant for  $x_2(t)<x<x_4(t)$; and $C_5C_6 e^{2F_1+2F_2}$ is dominant for $x_4(t)<x$. Then, it follows that there is a line soliton moving with the speed $V_{l2} = 4a_2^2$ on the vertical line $x=x_2(t)$ and a parallel soliton moving with the speed $V_{l1} = 4a_1^2$ along $x=x_4(t)$---see frame (a) in Fig. \ref{f07}.
    
    \item Let $t_0 = \left(\ln(1+C_6^{-1}) - \ln{C_5}\right)/8a_1a_2(a_1+a_2)$. When  $0<t<t_0$, then $1$ is dominant for $x<x_1(t)$; $(1+C_6)e^{2F_1}$ is dominant for $x_1(t)<x<x_5(t)$; $e^{2F_2}$ is dominant for $x_5(t)<x<x_4(t)$; and $C_5C_6 e^{2F_1+2F_2}$ is dominant for $x_4(t)<x$. Then, there are line solitons moving along $x=x_1(t)$ and $x=x_4(t)$ both having speeds $V_l = 4a_1^2$, while along the line $x=x_5(t)$ there is a lump chain moving with the speed $V_c$---see frame (b) in Fig. \ref{f07}.
    
    \item When $t > t_0$, $1$ is dominant for $x<x_1(t)$; $(1+C_6)e^{2F_1}$ is dominant for $x_1(t)<x_3(t)$; and $C_5C_6 e^{2F_1+2F_2}$ is dominant for $x_3(t)<t$. Hence, there is a line soliton moving with the speed $V_{l1} = 4a_1^2$ along $x=x_1(t)$ and a line soliton moving with the speed $V_{l2} = 4a_2^2$ along $x=x_3(t)$  -- see frame (c) in Fig. \ref{f07}.
\end{enumerate}
\begin{figure}[h!]
\centering
\includegraphics[width=1.0\textwidth]{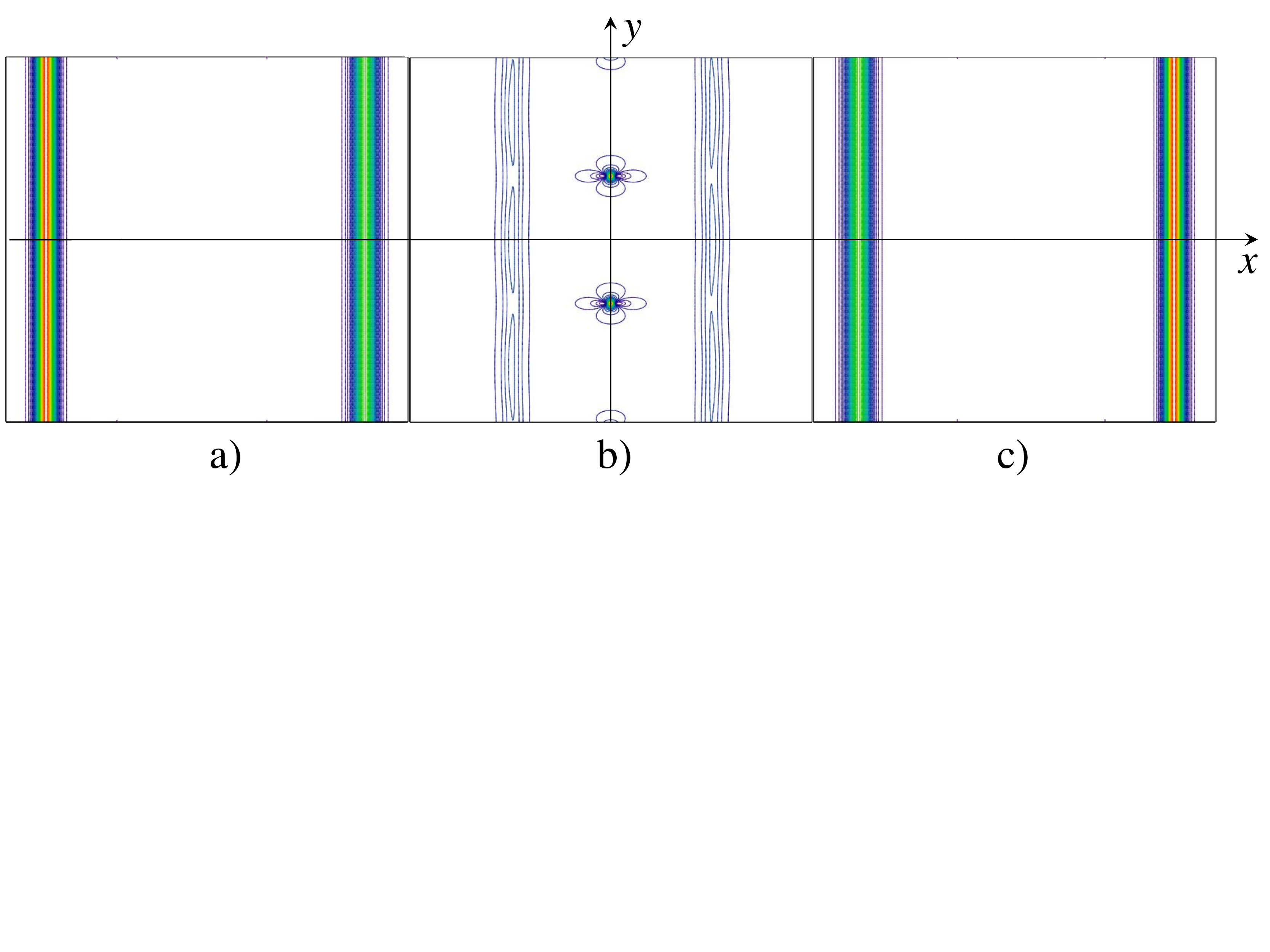}
\vspace{-7.cm}
\caption{Exchange type interaction of two line solitons with the parameters $a_1 = 0.4$, $a_2 = 0.5$ (the corresponding amplitudes are: $A_{e1} = 0.32$, $A_{e2} = 0.5$); lump amplitudes are $A_l = 3.2$. Frame (a), $t = -80$, $-90 \le x \le 10$; frame (b), $t = 16$, $-14.4 \le x \le 85.6$; frame (c), $t = 120$, $50 \le x \le 150$; in all frames $-100 \le y \le 100$.} %
\label{f07}
\end{figure}

Summarizing the described process of soliton interaction, we see that the solution has the following structure. A line soliton with the speed $V_{e2} = 4a_2^2$ at minus infinity approaches a slower moving soliton with the speed $V_{e1} = 4a_1^2$. At $t=0$, the left soliton emits a lump chain and slows down to $V_{e1} = 4a_1^2$. The lump chain propagates with the speed $V_c$ and at $t=t_0$ is absorbed by the right soliton accelerating it up to the speed $V_{e2} =  4a_2^2$. The lifespan $t_0$ of the lump chain increases logarithmically as $C_6 \to 0$. In this limiting case, solution based on the $\tau$-function (\ref{Eq.45}) reduces to the solution (\ref{Eq.39}) describing radiation of a lump chain by a line soliton.

Solution describing absorption of a lump chain by a line soliton and based on the $\tau$-function (\ref{eq:C4}) can be obtained in another limiting case. To this end, we present the $\tau$-function (\ref{Eq.45}) in the form:
$$
\tau(x,y,t) = C_6\left[\frac{1}{C_6}\left(1 + e^{2F_1} + e^{2F_2}\right) + e^{2F_1} + \frac{C_4}{C_6}e^{F_1+F_2}\cos{\left[\left(a_1^2-a_2^2\right)y\right]} + C_5  e^{2F_1+2F_2}\right]
$$
assuming that the coordinate $x$ in functions $F_1(x,t)$ and $F_2(x,t)$ is shifted to the right by an arbitrary value $x_0$. 
The constant factor $C_6$ can be omitted as it does not contribute to the solution in terms of $u(x,y,t)$. Consider then the limit when $\rho_3 \to \infty$, $\rho_2 \to -\infty$, $\rho_1 \to \infty$, and $x_0 \to -\infty$ such that $\rho_1 = \rho_3 - p_1 + p_2$, $\rho_2 = a_2(-\rho_3 + 2p_1 - p_2)/a_1 - p_3$, and $x_0 = (-\rho_3 + 2p_1 - p_2)/a_1$, where $p_1$, $p_2$, and $p_3$ are some constants. Then, $C_6 \to \infty$, and $\tau$-function reduces to Eq. (\ref{eq:C4}).

Note that the described process of exchange-type interaction of two line solitons is very similar to the interaction of two line solitons within the KdV equation when the ratio of their amplitudes at infinity $A_1/A_2 < 2.62$. Within the KP1 equation both these processes can occur; the KdV-type interaction occurs when two line solitons are unperturbed at infinity, whereas the KP-type interaction by means of a lump chain is, apparently, a special case when one of the line solitons has a specific infinitesimal modulation along its front. There is, however, one important feature that demonstrates a big difference in the interaction of line solitons of the KdV-type and KP-type. It is well-known (see, e.g., \cite{Ablowitz-1981}) that KdV solitons experience a phase shift after interaction, and the phase shift of each soliton is determined entirely by the spectral parameters $a_1$ and $a_2$:
\begin{equation}
\label{Eq.46}
\left(\Delta x_{KdV}\right)_{1,2} = \frac{1}{a_{1,2}}\ln{\left|\frac{a_1-a_2}{a_1+a_2}\right|}.
\end{equation}
However, when two KP-1 line solitons exchange a lump chain, the phase shift is determined not only by the spectral parameters but also by the parameters $\rho_1$ and $\rho_3$:
\begin{equation}
\label{Eq.47}
\left(\Delta x_{KP}\right)_{1,2} = \left(\Delta x_{KdV}\right)_{1,2} - \frac{1}{2a_{1,2}}\ln{\left[1 + e^{2(\rho_1 - \rho_3)}\right]}.
\end{equation}
This phase shift may be arbitrarily large. By analyzing the phase shift at plus and minus infinity, we are able to recognize whether the KP-1 solitons interacted according to the KdV approximation, or if they exchanged a lump chain. 

\end{enumerate}

\section{CONCLUSIONS}

Thus, in this paper, we have described the elementary acts of interactions of line solitons with lumps and with each other by means of lumps within the framework of KP1 equation. Such interactions are impossible within the framework of the KP2 equation applicable to media with the negative dispersion. Our description is based on the presentation of solutions in terms of the $\tau$-function and Grammian. We have presented a lump emission and absorption by a line soliton, interaction of a lump and a line soliton, resonant interaction of line solitons through a lump, and emission and absorption of a periodic chain of lumps by a line soliton. In a similar way, one can study more complex dynamics of a number of line solitons and lumps; some results in this direction have been obtained in Refs. \cite{Rao-2021, Guo-2021}. 

Note that, the apparent, ``sudden'' arising of a lump between two line solitons and it subsequent disappearing after absorption by the second line soliton, can be treated formally as the rogue wave formation. Indeed, in our variables, lump amplitude is eight times greater than the amplitude of a plane soliton, wheres according to the accepted criterion \cite{Kharif-2009}, the rogue wave is such a wave whose amplitude is two or more times greater than the average amplitude of background waves. 

In the conclusion, it is worth noting that the KP1 equation has infinitely many integrals of motion, although the set of such integrals, apparently, is incomplete \cite{Zakharov-2009}.
The first integrals of this set (mass, momentum, energy) play an important role in physical applications. 
The simplest in this set is the mass-conservation integral over the whole $x,y$-plane:
\[I_m = \iint\limits_{D} u(x,y,t)\,dx\,dy.\]
This improper double integral does not satisfy the condition of the Fubini theorem for the lump solution (\ref{Lump}) because function $u(x,y,t)$ does not vanish sufficiently rapidly when $x^2 + y^2 \to \infty$. 
Therefore, the result of integration depends on the order of integration over $x$ and $y$.
The higher-order integrals, including integral of energy and Hamiltonian, containing $u(x,y,t)$ in degrees higher than 2 are well-defined and can be evaluated with the help of the Fubini theorem.

Since this paper was submitted to the journal on 28.07.2021, some more papers containing similar results have been published \cite{Zhang-2022, Zhao-2022}. \\

Y.A.S. acknowledges the funding of this study provided by the Council of the grants of the President of the Russian Federation for the state support of Leading Scientific Schools of the Russian Federation (grant No. NSH-70.2022.1.5). V.E.Z. acknowledges the funding of this study provided by Russian Science Foundation (grant No. 19-72-30028).\\

\end{document}